\documentclass{article}
\usepackage{graphicx} % Required for inserting images
\usepackage{algorithm}
\usepackage{algpseudocode}
\usepackage{amssymb}
\usepackage{amsmath}
\usepackage{cite}
\usepackage{tikz}
\usetikzlibrary{quantikz2}
\usepackage{authblk}

\DeclareMathOperator*{\argminA}{arg\,min}

\def\BibTeX{{\rm B\kern-.05em{\sc i\kern-.025em b}\kern-.08em
    T\kern-.1667em\lower.7ex\hbox{E}\kern-.125emX}}

\begin{document}

% % Outcomment only when entries are known. Otherwise leave as is and 
% %   default values will be used.
% % \setcounter{page}{1}
% \VOLUME{35}%
% \NO{3}%
% % \MONTH{Xxxxx}% (month or a similar seasonal id)
% \YEAR{2024}% e.g., 2005
% % \FIRSTPAGE{000}%
% % \LASTPAGE{000}%
% \SHORTYEAR{24}% shortened year (two-digit)
% \ISSUE{0000} %

% \TITLE{Enhancing Quantum Optimization with Parity Network Synthesis}
% \author{Colin Campbell, Denny Dahl}
% \ARTICLEAUTHORS{%
% \AUTHOR{Colin Campbell}
% \AFF{Infleqtion, \EMAIL{colin.campbell@infleqtion.com}, \URL{}}
% \AUTHOR{Edward D Dahl}
% \AFF{IONQ, \EMAIL{denny.dahl@ionq.co}, \URL{}}
% }

% \ABSTRACT{
%     This paper examines QAOA in the context of parity network synthesis. We propose a pair of algorithms for parity network synthesis and linear circuit inversion. Together, these algorithms can build the diagonal component of the QAOA circuit, generally the most expensive in terms of two qubit gates. We compare the CNOT count of our strategy to off the shelf compiler tools for random, full, and graph-based optimization problems and find that ours outperforms the alternatives. }

% \KEYWORDS{quantum computing, compilation, combinatorial optimization}
\title{Enhancing Quantum Optimization with Parity Network Synthesis}
\author[1]{Colin Campbell}
\author[2]{Edward D Dahl}
\affil[1]{\textit{Infleqtion, Chicago, IL 60622, USA}}
\affil[2]{\textit{IonQ, Inc. 4505 Campus Dr. College Park, MD 20740}}
\date{}

\maketitle
\begin{abstract}
    This paper examines QAOA in the context of parity network synthesis. We propose a pair of algorithms for parity network synthesis and linear circuit inversion. Together, these algorithms can build the diagonal component of the QAOA circuit, generally the most expensive in terms of two qubit gates. We compare the CNOT count of our strategy to off-the-shelf compiler tools for random, full, and graph-based optimization problems and find that ours outperforms the alternatives.
\end{abstract}

\section{Introduction}

Combinatorial optimization is a computationally difficult class of problems that arises naturally in many fields 
% [\textcolor{red}{missing ref}]
. The Quantum Approximate Optimization Algorithm (QAOA) \cite{farhi} has been of great interest to researchers looking to solve optimization problems on a quantum computer. Since its initial proposal, variations on the original formulation have cultivated a better understanding of the strengths and weaknesses of the algorithm \cite{Bravyi_rqaoa, Hadfield_alternating_operator, Tomesh_2022, Shaydulin_2019, zhu2022adaptive}.
While the many variants of QAOA change the structure of the original circuit, the main kernel of the algorithm is usually the application of a parameterized diagonal unitary generated from the optimization problem, denoted $U_C(\gamma)$. Typically in QAOA, $U_C(\gamma)$ accounts for all or most of the two qubit gates in the circuit. Often, work on QAOA focuses on quadratic problems for which the two qubit cost of $U_C(\gamma)$ cannot obviously be reduced; however, considering the broader space of higher order binary optimization problems opens the door for more powerful techniques in optimizing gate count and brings applications within striking distance of real hardware implementations. Demonstrating advantage with QAOA can be challenging. In some cases QAOA is not better than classical algorithms \cite{Marwaha_maxcut_loses}. Even without a strict proof of advantage, the algorithm remains popular, and serves as a useful tool for understanding how to utilize quantum resources effectively for optimization problems. This paper offers a technique for constructing the diagonal unitary $U_C(\gamma)$ based on the principles of parity network synthesis, but it should be noted that techniques for building diagonal unitaries can be useful in other contexts \cite{Cowtan_UCC}. We present two greedy algorithms that, in combination, efficiently construct circuits of this type for sparse optimization problems and provide numerical evidence demonstrating that the proposed technique outperforms alternative compilation strategies by comparing the CNOT count for several families of optimization problems. 

During the drafting of this manuscript, the authors found related work in \cite{Vandaele_phase_polynomials} and \cite{De_Brugi_circuit_inversion} that independently propose similar algorithms and evaluate their performance on random problems. In this new work, the authors corroborate the evidence of performance increase for random problems and look at applications in more targeted contexts such as bounded order problems and graph coloring. We also focus on comparing against ``out of the box" compilation software by using Qiskit \cite{Qiskit} to establish a baseline in our analysis. The authors have done their best to state where overlap occurs throughout the rest of the paper.

\section{Problem Formulation}
% There are two common ways of formulating combinatorial optimization problems. One which is common in operations research is to use variables $x_i \in \{0,1\}$, while the other, which is popular with physicists is to use variables $s_i \in \{1, -1\}$. The relationship between these variables is a straightforward algebraic expression:
% \begin{equation*}
%     s_i = 1 - 2x_i
% \end{equation*}

% In this paper we will be using $\{0, 1\}$ to describe the process of building the diagonal unitary. In particular we will be treating bit strings often written as $b \in \{0, 1\}^n$ as vectors in the vector space $\mathbb{F}_2^n$. Due to the relationship between $s$ and $x$ variables above,
In studies of QAOA, there is a well known relationship between combinatorial optimization problems typically posed as
\begin{equation}
    \argminA_{x\in \{0, 1\}^n}x^T Q x
\end{equation}
and finding the ground state of an Ising Hamiltonian of the form
\begin{equation}
    H = \sum_i b_iZ_i + \sum_{i<j}b_{ij}Z_iZ_j
\end{equation}
where $Z_i$ is the Pauli $Z$ operator with eigenvalues in $\{1, -1\}$ operating on qubit $i$. For quadratic optimization problems, $Q$ is a matrix and represents a polynomial $f(x) = \sum_iq_ix_i + \sum_{i<j}q_{ij}x_ix_j$. This function takes bit strings of length $n$ and maps them to real values and is known as a psuedo-boolean function. For optimization problems of higher order than quadratic, we generalize both the quadratic polynomial $f$ and the second order Ising model $H$ to allow for terms of arbitrary order, using $\mathbb{F}_2^n$ to denote the vector space of bit strings since it is standard when discussing parity networks and is helpful in understanding the motivation for the greedy algorithm.
% For a pseudo-boolean function $f:\mathbb{F}_2^n \to \mathbb{R}$, any binary optimization problem of the form
% \begin{align*}
%     \mathrm{arg min}_{x\in \mathbb{F}_2^n} f(x)
% \end{align*}
% is equivalent to finding the ground state of an Ising model of the following form:
Generalizing the second order Ising model yields
\begin{equation}
    H = \sum_{i_1} b_{i_1}Z_{i_1} + \sum_{i_1 < i_2} b_{i_1, i_2} Z_{i_1}Z_{i_2} + ... + \sum_{i_1< ...< i_n}b_{i_1,...,i_n}Z_{i_1}...Z_{i_n}
\end{equation} and generalizing the quadratic polynomial $f$ yields
\begin{equation}
    f(x) = \sum_{y\in\mathbb{F}_2^n}\hat{f}(y)(x_1y_1\oplus\ x_2y_2\oplus...\oplus x_ny_n)
\end{equation}
% where $\chi_y(x) = x_1y_1\oplus\ x_2y_2\oplus...\oplus x_ny_n$ and 
where $\hat{f}(y)$ plays the same role of term coeficients as the $b_{i_1,...i_k}$ in (3). See \cite{Amy_gray_synth, BOROS2002155} for more thorough treatment of psuedo-boolean functions.
Limiting the terms to second order, Hamiltonian (3) is the well-known Ising model. The generalized version is related to the ``infinite range" spin glass model, which admits a known solution \cite{infinite_range_spin_glass}; however, for coefficients produced from an optimization problem, the assumption of independently distributed interactions does not hold, making this problem suitably hard.
In fact, if all weights are nonzero, the Hamiltonian (psuedo-boolean function) has exponentially many terms, making writing it down is as difficult as finding its lowest energy states via exhaustive enumeration. To avoid this situation, an important property for choosing problems for QAOA is sparsity. A problem is sparse if as the number of variables in the problem grows, the number of terms in the Hamiltonian is bounded by a polynomial of fixed degree. 
This constraint is well-justified because many optimization problems are limited to a specific order $k$ that does not grow with $n$. For instance, MaxCut is bounded by degree 2 and 3SAT is bounded by degree three. Quadratic optimization problems like MaxCut are a good choice in terms of sparsity because they are difficult to solve classically while having only a quadratic number of terms, however, recent work has also explored the benefits of encoding problems over higher order terms\cite{qaoa_of_the_highest_order, Pelofske_higher_order_google, glos2020spaceefficient, Kapit_3sat}. While higher order problems can be reduced, the overhead from ancilla and slack variables can increase the qubit count, gate count, and difficulty of the quantum-classical optimization loop.

Given a sparse $H$ as defined above and the basis gate set of $\{CNOT, R_z\}$, the task is to implement $U=e^{i\gamma H}$ using as few CNOT gates as possible. \cite{Amy_gray_synth, Vandaele_phase_polynomials, De_Brugi_circuit_inversion} contain strategies for completing this task. This work aligns most closely with \cite{Vandaele_phase_polynomials}.

First we discuss an approach to building this circuit that is common when constructing QAOA circuits of low order Hamiltonians.
Since all the terms in $H$ commute with each other, the unitary $e^{i\gamma H}$ can be broken into a product of commuting operators
\begin{equation}
    e^{i\gamma H} = \prod_{k=1}^n\left( \prod_{i_1<...<i_k} e^{i\gamma b_{i_1...i_k} Z_{i_1}...Z_{i_k}} \right).
\end{equation}
Each term in the innermost product is called a ``phase gadget" \cite{phase_gadget_synthesis}. Building a circuit to implement a phase gadget can be done recursively using the following definition:
\begin{align*}
    \Phi_1(\alpha) \coloneqq R_z(\alpha) \quad \Phi_{n+1}(\alpha) \coloneqq \left(\text{CNOT}\otimes 1_{n-1}\right) \left(1_1 \otimes \Phi_n(\alpha)\right) \left(\text{CNOT}\otimes 1_{n-1}\right).
\end{align*}

% Given a fixed $k$,  Exponentiating the Hamiltonian has the following result:
% \begin{equation}
%     U = e^{i\gamma H} = \exp\left(i\gamma \sum_ib_iZ_i\right) = \bigotimes_iR_z(\gamma b_i)
% \end{equation}

% The original sum breaks into a product because the $Z_i$'s commute, and for higher order cases this commutation relation will still hold. Therefore one option for decomposing $U$ is to use a product of diagonal gates
% \begin{equation*}
%     \prod_{i_k}\exp\left(i\gamma b_{i_1...i_k}Z_{1_1}...Z_{i_k}\right)
% \end{equation*}.
% These gates are often called ``phase gadgets" [cqc reference] where the phase gadget is defined recursively as follows:
% [the recursive definition of the phase gadget]
There are two key observations to make from this recursive definition. First, each term in the original Hamiltonian can be implemented with $2(k-1)$ CNOT gates, where $k$ is the order of the term. Since each term can be built independently from the rest, we can establish an upper bound for the number of two qubit gates in the circuit implementing $e^{i\gamma H}$. Let $S_k = \{b_{i_1,...i_k}\neq 0\}$ be the nonzero weights of order $k$. The number of CNOT gates needed to implement the circuit is not greater than
\begin{equation}
    \sum_{k=2}^{n}2(k-1)|S_k|.
\end{equation}
The second key observation is that building the circuit recursively and choosing the order of the phase gadgets results in a large number of equivalent circuits with different opportunities for cancellation, as illustrated in Figure 1.

\begin{figure}[t]
    \centering
    \begin{quantikz}
        & \ctrl{2} & & & & \ctrl{2} & \ctrl{1} & & \ctrl{1} \\
        & & \ctrl{1} & & \ctrl{1} & & \targ{} & \gate{R_z} & \targ{}\\
        & \targ{} & \targ{} & \gate{R_z} & \targ{} & \targ{} & & & \\
    \end{quantikz} \\
    \centering
    \begin{quantikz}
        & \ctrl{1} & & & & 
        \ctrl{1}\gategroup[2,steps=2,style={inner
        sep=6pt, dashed}]{} & \ctrl{1} & & \ctrl{1} & \\
        & \targ{} & \ctrl{1} & & \ctrl{1} & \targ{} & \targ{} &
        \gate{R_z} & \targ{} & \\
        & & \targ{} & \gate{R_z} & \targ{} & & & & &
    \end{quantikz}
\    \caption{While the first circuit and the second circuit produce identical unitary operators, the second has an obvious opportunity for cancellation because a CNOT gate in the second phase gadget cancels with one in the first phase gadget, demonstrating how different recursive decompositions can have different opportunities of reducing gate count.}
\end{figure}
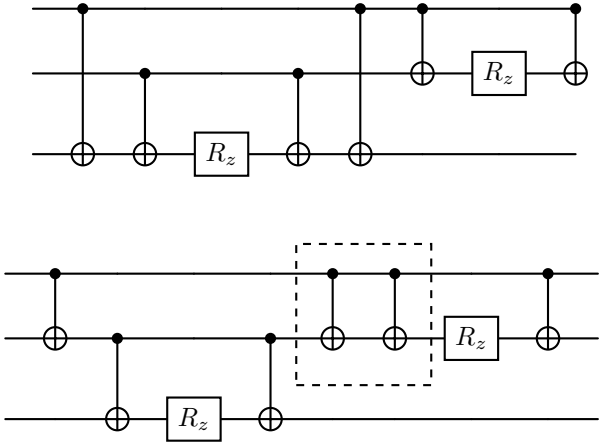

Building each phase gadget individually to create the final circuit is a specific example of a more general family of circuits called parity networks. When performing CNOT gates between qubit wires, one can keep track of the parities of those wires by annotating them in the circuit. Parities written as $x_1\oplus x_2 \oplus ...$ map easily to elements of $\mathbb{F}_2^n$. For example, in a five qubit system $x_0 \oplus x_2 \oplus x_3$ would map to $10110$.
For input $S\subseteq \mathbb{F}_2^n, A\in \text{GL}(n,\mathbb{F}_2)$, a parity network is a circuit of only CNOT gates for which every element of $S$ appears in the annotated circuit and the final state of the wires is $A$. Figure 2 gives an example of a parity network for $S = \{00010, 00100, 00101, 01111, 11010, 11011\}$, $A=I$. For the purposes of QAOA, $A$ will always be the identity, or a permutation thereof, due to the mixer that follows $U_C(\gamma)$.
The final step to complete the unitary is to insert $R_z(b_{i_1...i_k}\gamma)$ gates in the parity network where parities on the qubit wires correspond to the weights in the problem Hamiltonian.
For a more thorough treatment of parity networks in the context of phase polynomials, see \cite{Amy_gray_synth, Vandaele_phase_polynomials}.
It has been shown that for limited conditions on the coefficients in the phase gadgets, the CNOT-minimal parity network contains the minimum number of CNOTs to implement the overall unitary and that there exists an algorithm called gray-synth that is optimal in the case where the parity network contains all possible parities \cite{Amy_gray_synth}.
For the purposes of the diagonal portion of QAOA, just having the parity network is not enough to get the number of CNOT gates in the circuit. After building the parity network, one can use a method to invert the parities like that in \cite{patel2003efficient}. The gate cost of the parity network is generally much lower because it does not to do each inversion individually as would be the case in the recursive strategy for building the circuit.

\begin{figure}
\centering
    % \begin{subfigure}
        \begin{quantikz}
            \lstick{$x_0$} & & & & & \ctrl{2} & & & \ctrl{3} & & & & \ctrl{3} & \ctrl{2} & &\\
            \lstick{$x_1$} & & \ctrl{2} & &  & & & &  & & &  & &  &\ctrl{2} & \\
            \lstick{$x_2$} & \ctrl{2} & & & & \targ{} & \wire[l][1]["x_0\oplus x_2"{above,pos=0.2}]{a} & \ctrl{2} & &&& & & \targ{} & \wire[l][1]["x_2"{above,pos=.6}]{a} & \\
            \lstick{$x_3$} &  & \targ{} & \wire[l][1]["x_1\oplus x_3"{above,pos=0.2}]{a} & \ctrl{1} & & & & \targ{} & \wire[l][1]["x_0\oplus x_1 \oplus x_3"{above,pos=-.3}]{a} & &\ctrl{1} & \targ{} & \wire[l][1]["x_3\oplus x_1"{above,pos=.4}]{a} & \targ{} & \wire[l][1]["x_3"{above,pos=.5}]{a} \\
            \lstick{$x_4$} & \targ{} & \wire[l][1]["x_2\oplus x_4"{above,pos=0.2}]{a} & & \targ{} & \wire[l][1]["x_1\oplus x_2\oplus x_3\oplus x_4"{above,pos=-.35}]{a} & & \targ{} & & \wire[l][1]["x_0\oplus x_1\oplus x_3\oplus x_4"{above,pos=.5}]{a} & & \targ{} & \wire[l][1]["x_4"{above,pos=0.6}]{a} & & &
        \end{quantikz}
    % \end{subfigure}
    % \begin{subfigure}
        \begin{quantikz}
            \lstick{$x_0$} & & & & & & \ctrl{2} & & \ctrl{3} & & & \ctrl{3} & \ctrl{2} & & \\
            \lstick{$x_1$} & & & \ctrl{2} & &  & & & & & & & & \ctrl{2} & \\
            \lstick{$x_2$} & \gate{R_z} & \ctrl{2} & & & & \targ{} & \ctrl{2} & & & & & \targ{} & & \\
            \lstick{$x_3$} & \gate{R_z} & & \targ{} &  & \ctrl{1} & &  & \targ{} & \gate{R_z} & \ctrl{1} & \targ{} & & \targ{} & \\
            \lstick{$x_4$} & & \targ{} & \gate{R_z} & & \targ{} & \gate{R_z} & \targ{} & \gate{R_z} & & \targ{} & & & &
        \end{quantikz}
    % \end{subfigure}
    \caption{The first circuit shows a parity network over $S = \{00010, 00100, 00101, 01111, 11010, 11011\}$ with the parities on each wire displayed as they are changed by CNOT gates. The final parities have been made to be the same as the initial parities. The second circuit includes the $R_z$ gates at the spots where the parity network has parities in $S$ to build the full unitary from the parity network. The angles in the $R_z$ gates are proportional to the weights in equation (5) and the $\gamma$ parameter in QAOA}
    \label{fig:enter-label}
\end{figure}
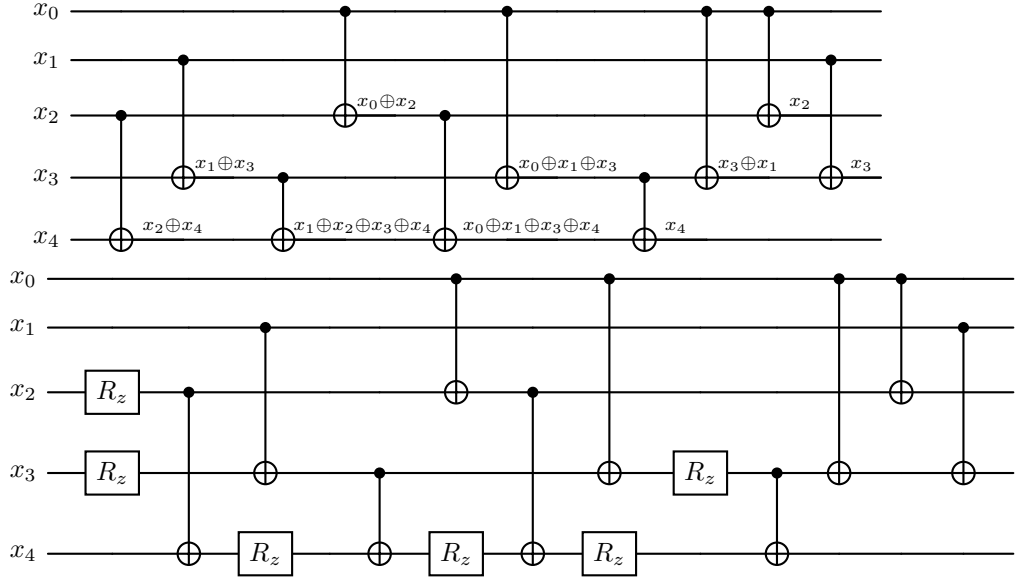

\section{A Greedy Algorithm}

In a parity network, each wire corresponds to a vector in $\mathbb{F}_2^n$, and the state of all $n$ wires corresponds to a matrix in $\mathbb{F}_2^{n\times n}$.
In Figure 2 the first CNOT gate brings the $x_4$ wire from $00001$ to $00101$.
Because a parity network only includes CNOT gates, any change in a wire's state is reversible (by applying the same CNOT gate).
Therefore a snapshot of the wires at any point in the circuit is a basis for $\mathbb{F}_2^n$.
The wires start in the standard basis of bit strings with unit Hamming weight. 
% $\{e_i\} = \{x \in \mathbb{F}_2^n:|x|=1\}$. 
Each CNOT gate in the circuit replaces the parity of the target with the bit-wise the sum of the control and the target. Therefore, if a parity in $S$ appears in the state of the qubit wires, it also appears in the current basis. The parity network in conjunction with the return journey for the exponentiated Hamiltonian circuit can be thought of as a sequence of basis transformations starting from the standard basis and returning to the standard basis, such that all elements of $S$ appear in at least one basis in the sequence. Building the parity network and returning to the standard basis can be thought of as two separate components, which is why we present them as separate algorithms.
The recursive phase gadget construction does both at once, effectively building a parity network and returning to the standard basis for each element in $S$.
This inherent backtracking makes the procedure very inefficient compared to building the parity network all at once and then returning to the standard basis at the end.

The number of CNOT gates needed to include any new parity in another basis is equal to the Hamming weight of that parity in the current basis. The greedy algorithm chooses the closest state by Hamming weight, applies a CNOT gate, updates the remaining parities to be in terms of the new basis, and removes any with Hamming weight one.
Re-writing the current set of target parities in the current basis is relatively efficient. When $\text{CNOT}_{i,j}$ takes a basis state $x_j$ to $x_j\oplus x_i$, for each parity $s$, replace $s_i$ with $s_i\oplus s_j$.
If after this replacement procedure a parity has unit Hamming weight, that parity has been included in the current basis and can be removed from the set of remaining targets.
Algorithm 1 gives a pseudocode implementation of this process, similar to\cite{Vandaele_phase_polynomials}. After synthesizing the parity network, the return to the identity can be thought of as inverting the matrix representing the current state of the qubit wires. \cite{patel2003efficient} contains an asymptotically optimal method for performing this inversion, but in our algorithm we take advantage of the fact that Algorithm 1 produces triangular matrices to implement a greedy version of Gaussian elimination like\cite{De_Brugi_circuit_inversion}.

\begin{algorithm}
\caption{Greedy Parity Network Synthesis}
\textbf{Input: $S\subseteq \mathbb{F}_2^n$} \\
\textbf{Output: $U \in \text{SU}(2^n)$}
\begin{algorithmic}[1]
\State $U \gets I$
\While{$S$}
    \State $y = \argminA_{s\in S}|s|$ \Comment{Choose parity with the smallest Hamming weight}
    \State $I \gets \{1\leq i \leq n | y_i = 1\}$ \Comment{Find potential qubits for CNOT}
    \State $i \gets \min(I)$ \Comment{Smallest as contol}
    \State $j \gets \min(I \setminus \{i\})$ \Comment{Next smallest as target}
    \State $U \gets \text{CNOT}_{i,j}U$
    \State $S' \gets \{\}$
    \For{$s \in S$}
        \State $s' \gets s$
        \State $s'_i = s_i \oplus s_j$ \Comment{Update parity a bit position $i$}
        \If{$|s'| > 1$} \Comment{Target parity has been reached}
            \State $S' \gets S' \cup \{s'\}$
        \EndIf
    \EndFor
    \State $S \gets S'$
\EndWhile
\State \Return $U$
\end{algorithmic}
\end{algorithm}

\begin{algorithm}
\caption{Greedy Gaussian Elimination}
\textbf{Input: $A \in \mathbb{F}_2^{n\times n}$} \\
\textbf{Output: $U$}
\begin{algorithmic}[1]
    \State $U \gets I$
    \While{$|A|_\infty > 1$} \Comment{If $A\neq I$, the maximum row sum is greater than 1}
        \State $C \gets \{(i,j) \in [1,...,n] \times [1,...,n] : i<j\}$ \Comment{Combinations of row indices of $A$}
        \State $l, m \gets \max_{i, j \in C} \left( \max\left( |A_i|, |A_j|\right) - |A_i\oplus A_j|\right)$  \Comment{Best potential row operation}
        \If{$|A_l|<|A_m|$} \Comment{Row with smaller Hamming weight is control}
    \State $U \gets \text{CNOT}_{l,m}$
    \State $A_m \gets A_l \oplus A_m$
\Else
    \State $U \gets \text{CNOT}_{m,l}$
    \State $A_l \gets A_l \oplus A_m$
\EndIf 
    \EndWhile
\State \Return $U$
\end{algorithmic}
\end{algorithm}

\section{Evaluating Performance}

In this section we compare the CNOT count of using Algorithms 1 and 2 with two other methods available in the quantum programming toolkit Qiskit: the default compiler at the highest optimization level and the built-in implementation of gray-synth. Since Qiskit uses \cite{patel2003efficient} to perform the return journey after building the parity network, we also compare against gray-synth using Algorithm 2 to perform the return journey. Where applicable we also plot the upper bound in (6).
To demonstrate that Algorithms 1 and 2 as a circuit building technique work on different types of problems, we consider three families of problems: random problems of a fixed size, full problems up to order $k$, and graph coloring on connected caveman graphs.
% To explore a compelling array of sparse problems we compare the CNOT gate count of each technique on four families of problems: random problems of fixed size, full kth order problems from k=3 to k=5, graph coloring on connected caveman graphs, and graph coloring on connected erdos-renyi graphs for several values of $p$.

Figure 3 displays the results for random problems over $n$ variables of size 100. Of the $2^n-1$ possible parities, 100 were chosen uniformly at random without replacement. For cases where 100 is greater than the number of possible parities, this sampling technique simplifies to the parity network over all parities.
This process was then repeated 30 times. Figure 3 plots the average and one standard deviation from the mean for each set of 30 samples. 
By construction, this family of problems is sparse because the number of terms is constant as the problem size grows. While random, this family of problems is not without some structure.
The average term order is $n/2$ because when sampling uniformly at random there is an independent $\frac{1}{2}$ chance for each bit to be on.
At small values of $n$, gray-synth performs well because every parity is included in the network.
As $n$ grows, both gray-synth and greedy perform well compared to the Qiskit compiler because Qiskit is building phase gadgets individually, resulting in so-called ``ladder" structures which track the upper bound in (6). We plot the ladder method explicitly in Figure 3 to support this conclusion.
% As $n$ grows however, both gray and greedy perform much better than standard Qiskit. The difference in performance comes from Qiskit using the ``ladder" method, which tracks the upper bound calculated in (equation).
Between the two well-performing methods, greedy is competitive at low $n$ and achieves better scaling as $n$ increases.
This better scaling behavior for random problems corroborates the results in \cite{Vandaele_phase_polynomials}. Although random problems are a good starting metric, optimization problems found out in the real world generally do not fall nicely into this category and will have some structure that can be exploited. The next example considers a much more structured family of problems that is more relevant to general problems of interest.

\begin{figure}[t]
    \centering
    \includegraphics[width=0.5\textwidth]{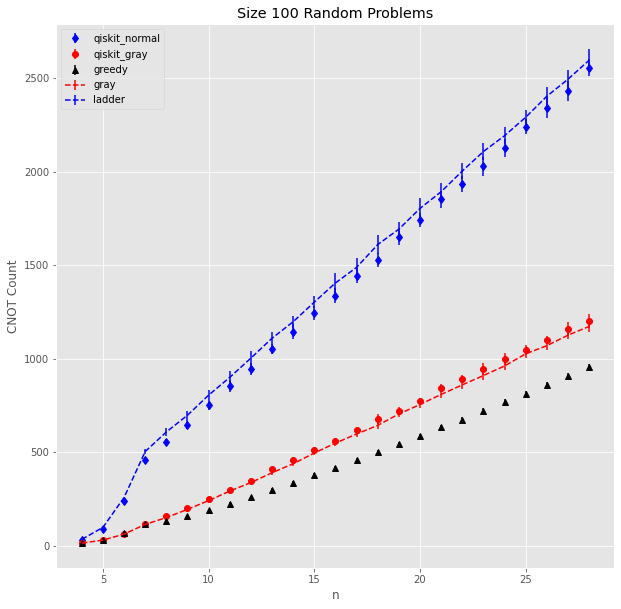}
    \caption{Random parity networks of size 100 averaged over 30 samples. Also included is the result of our implementation of gray-synth to build the parity network and Algorithm 2 to perform the return journey. There is not much difference in CNOT count, indicating that the return process generally does not cost many CNOT gates compared to building the parity network.}
    % \label{fig:2l}
\end{figure}

Often it is useful to consider optimization problems that include all parities up to a particular order. For instance, quadratic optimization problems are a specific choice of $k=2$.
Expressing problems where correlations between small sets of variables can be useful for space-efficient encodings of problems like the traveling salesperson problem \cite{glos2020spaceefficient} and graph coloring \cite{qaoa_of_the_highest_order}.
These problems are extremely versatile in their potential applications because they are so general. Figure 4 compares the CNOT cost between techniques for $k=3,4,5$.
Once again, the greedy algorithm performs the best, but this time there is not a universal winner between the Qiskit compiler and gray-synth. One reason to expect the greedy algorithm to perform well on this problem is that the target parities are always close to the current basis, whichever basis that may be. Being close to all the target parities prevents the greedy algorithm from running into the trap of choosing an immediately close parity at the cost of moving far from the rest. The results from this type of problem suggest that the greedy algorithm performs well on small extensions of the quadratic problem formulation, formulations that include more higher order terms that are relatively low in order.
\begin{figure}
    \centering
    \includegraphics[width=\textwidth]{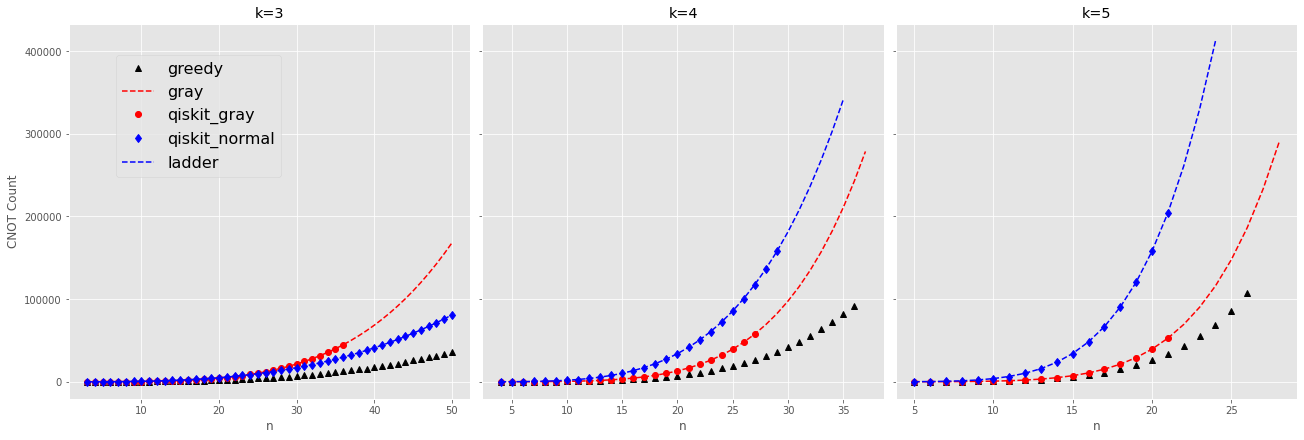}
    \caption{Between gray-synth, the Qiskit, compiler, and greedy parity network synthesis, the gate savings becomes more apparent as the order increases. Due to the limits on time and computational resources, Qiskit's gray-synth implementation could not be run as long as the other techniques, but the difference in gate count between our implementation with greedy Gaussian elimination and Qiskit's implementation appear negligible.}
    \label{fig:2}
\end{figure}
% while qiskit's implementation of gray-synth performs much worse in terms of the total number of CNOT gates. (For reference, we also plot a lower bound using equation (that one equation). Although this lower bound is not strict, we see that the greedy algorithm tracks it fairly closely, implying that our strategy is not just an attractive alternative to other methods for this problem, but a good strategy to use in a vaccuum)
\begin{figure}[!ht]
    \centering
     \includegraphics[width=0.55\textwidth]{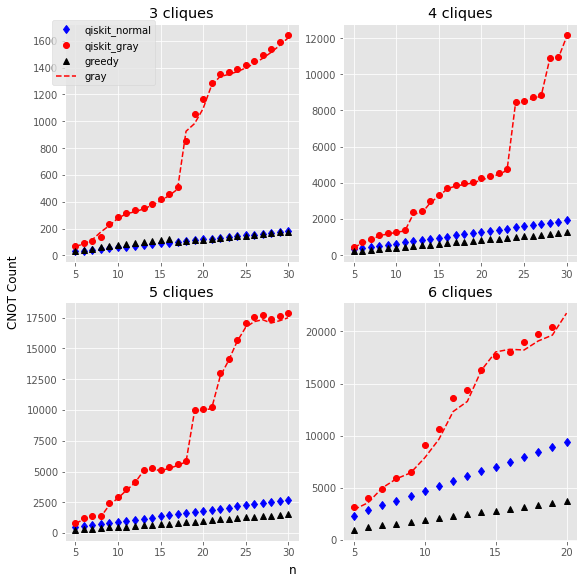}
     \caption{The CNOT cost of the circuits for graph coloring on connected caveman graphs increases with both the number of cliques and the clique size. In this case gray-synth performs the worst, while Qiskit's compiler tracks closely with the greedy algorithm until it becomes less competitive for higher clique sizes.}
\end{figure}
Random problems and full problems are two ends of the spectrum of sparse problems. Random problems lack structure to exploit, while k-bit problems have a structure that is easy to exploit. Real optimization problems will be somewhere in the middle of the spectrum. In order to challenge the algorithm to be useful for problems that have some structure without being too dense, we consider graph coloring on connected caveman graphs. Connected caveman graphs are of interest because of their community structure\cite{Watts_community_graphs, Walsh_search_in_small_world} as well as their simplicity. These graphs have two parameters--the number of cliques $n$ and the size of each clique $k$. The graphs are created by building $n$ instances of $k$-complete graphs and removing two edges from each to connect them all together in a cycle. To turn these graphs into an optimization problem, we allow the graph $k-1$ colors and follow the prescription of \cite{qaoa_of_the_highest_order} to build the higher order optimization problem. This prescription associates decision variables to the bits of a binary number representing the color of each node in the graph. While the graph coloring problem on this family of graphs is not particularly difficult as a problem, investigating the gate cost is fruitful for investigating how the various techniques here perform on a problem with some real underlying community structure and a unique incorporation of higher order terms.
The scaling comparison of the techniques on this family of graphs is shown in Figure 5. The number of CNOT gates grows with both $n$ and $k$, but there is a clear discrepancy in performance. While the Qiskit compiler and the greedy algorithm are competitive, especially in the low number of cliques, gray-synth apparently struggles with this problem.

Together these examples indicate that for many families of optimization problems, using Algorithms 1 and 2 to build the QAOA circuit is a good first option. Whether the problem has almost no structure, is very structured, or falls somewhere in between, the greedy algorithm is a good way to build a circuit that is frugal in its CNOT resources for solving higher order optimization problems with QAOA.

\section{Conclusion}
This paper makes an addition to the subject of using QAOA to explore quantum optimization problems with higher order terms by proposing and evaluating a greedy technique for building the main component of QAOA. This technique is comprised of Algorithm 1 to build a parity network in a CNOT-frugal way and Algorithm 2, which completes the unitary by returning the parities to their initial states. We offer numerical evidence to show that for random problems, full problems of fixed order, and graph coloring problems the technique performs better than common alternatives. While there are problems for which alternatives like gray-synth perform well, we find that greedy parity network synthesis is a great choice for problems that exhibit sparsity in the terms included. We argue that sparsity is a good criteria to use for problems to consider for QAOA because they avoid the trap of needing exponentially many gates in the circuit or exponentially many terms in the Hamiltonian.

\bibliographystyle{unsrt}
\bibliography{references2}

% \begin{figure}
%     \centering
%     \includegraphics[width=\textwidth]{erdos-renyi_example_figure.png}
%     \caption{Caption}
%     \label{fig:4}
% \end{figure}

\end{document}